\documentclass[a4paper,UKenglish,cleveref, autoref, thm-restate, nolinenumbers]{lipics-v2021}

\pdfoutput=1 
\hideLIPIcs  

\title{Are Dependent Types in Set Theory Feasible?}

\newcommand{\mailto}[1]{#1}

\author{Yunsong Yang}{Laboratory for Automated Reasoning and Analysis, EPFL, Lausanne, Switzerland, \and KTH Royal Institute of Technology, Stockholm, Sweden \and \url{https://people.epfl.ch/yunsong.yang} }{\mailto{yunsong@kth.se}}{https://orcid.org/0009-0002-7654-3995}{}

\author{Simon Guilloud}{Laboratory for Automated Reasoning and Analysis, EPFL, Switzerland \and \url{https://people.epfl.ch/simon.guilloud} }{\mailto{simon.guilloud@epfl.ch}}{https://orcid.org/0000-0001-8179-7549}{}

\author{Viktor Kun\v{c}ak}{Laboratory for Automated Reasoning and Analysis, EPFL, Switzerland \and \url{https://lara.epfl.ch/~kuncak/} }{\mailto{viktor.kuncak@epfl.ch}}{https://orcid.org/0000-0001-7044-9522}{}

\authorrunning{Y. Yang, S. Guilloud and V. Kun\v{c}ak} 

\Copyright{Yunsong Yang, Simon Guilloud and Viktor Kun\v{c}ak} 



\category{Short paper} 

\relatedversion{} 

\supplement{Source code and mechanized proofs are available on Zenodo at \url{https://doi.org/10.5281/zenodo.18997443}}

\nolinenumbers 

\EventEditors{}
\EventNoEds{0}
\EventLongTitle{}
\EventShortTitle{}
\EventAcronym{}
\EventYear{}
\EventDate{}
\EventLocation{}
\EventLogo{}
\SeriesVolume{}
\ArticleNo{}

\newif\ifcomments
\commentstrue        

\usepackage{xcolor}
\usepackage{listings}

\lstdefinelanguage{Lean}{
  morekeywords={def,theorem,inductive,structure,universe,variable},
  sensitive=true,
  morecomment=[l]{--},
  morecomment=[s]{/-}{-/},
  morestring=[b]"
}

\lstdefinelanguage{Scala}{
  morekeywords={val,def,object,class,case,trait,fun},
  sensitive=true,
  morecomment=[l]{//},
  morecomment=[s]{/*}{*/},
  morestring=[b]"
}

\DeclareMathOperator{\Pii}{Pi}
\DeclareMathOperator{\abs}{abs}
\DeclareMathOperator{\app}{app}
\DeclareMathOperator{\universeOf}{universeOf}

\begin{document}

\maketitle

\begin{abstract}
Following the types-as-sets paradigm, we present a mechanized embedding of dependent function types with a hierarchy of universes into schematic first-order logic with equality, with axiom schemas of Tarski-Grothendieck set theory. We carry this embedding in the Lisa proof assistant. On top of this foundation, we implement a proof-producing bidirectional type-checking tactic to compute proofs for typing judgements, with partial support for subtyping. We present examples showing how our approach enables automated reasoning for dependent types that is fully verified from set-theoretic axioms and deduction rules for schematic first-order logic with equality. Because types are merely sets, the resulting formalism supports equality that applies to all types and values and permits the usual substitution rules.
\end{abstract}

\section{Introduction}
\label{sec:introduction}

Zermelo-Fraenkel set theory with the axiom of choice (ZFC, \cite{jechSetTheory1978}) has remained the standard foundation of mathematics for over a century. In contrast, the landscape of interactive theorem proving is dominated by systems based on dependent type theory (DTT), such as Rocq~\cite{barras1997coq} and Lean~4~\cite{moura2021lean}. While the foundations of these systems offer useful practical properties, such as built-in computation, they are typically more complex to implement and validate than first-order logic with ZFC axioms. This is attested, for example, by the size of the logical kernel of these tools, the complexity of their core deduction rules, and the number of implementation bugs that have been discovered in practice over the years~\cite{carneiro2025lean4leanverifyingtypecheckerlean}. 

The successes of DTT-inspired features---including predictable type checking, built-in evaluation, and simplified structural recursion---motivate the development of similar features within set-theory-based proof assistants. Such developments also represents a step towards system interoperability via the transfer of proofs between proof assistants. Proof transfer can occur either at a low level via the mechanized translation of kernel-level statements and proofs, or at a higher level by using, for example, increasingly effective LLMs \cite{urban2026130klinesformaltopology} to translate user-written high-level proof scripts. In either case, the receiving system should be able to simulate the key features of the host system faithfully and without complex encoding. We believe this work is timely especially in the light of popularity of Lean among mathematicians. This popularity is, in our experience, sometimes met with the questioning of the necessity of using type theory as underlying foundation.

We thus aim to develop dependent type theory as a soft type system for a set-theory-based proof assistant. Whereas soft type systems, including one for Mizar's set theory, were considered in the past \cite{10.1007/978-3-540-74591-4_28}, here we aim for an implemented solution targeting dependent types. For the present paper, we build on the recent version of Lisa~\cite{guilloud2023lisa}, a proof framework with a small logical kernel based on first-order logic and set theory, which has already been used to mechanize higher-order logic and simulate proofs from HOL Light~\cite{guilloud2024mechanized}. While the prior work supports only the simply typed lambda calculus with top-level polymorphism, in the present paper we embed \textit{dependent function types with universes}, which form the core of the calculus of constructions (CoC, \cite{coquandCalculusConstructions1988}); a key part of the foundation of Rocq and Lean. The resulting interface offers the syntax and key properties of dependent type theory, while ensuring that type checking and evaluation are formally justified using set-theoretic axioms. (We do not discuss inductive types or recursive function definitions in this paper; \cite{guilloud2024mechanized} presents an implementation usable in the simpler HOL settings.) While our encoding is similar in spirit to previous work constructing set-theoretic models of CiC, our objective is practical and focused on shallow, syntactic translation: for example, typing judgements are represented as first-order sequents, and not as sets (as in e.g.  \cite{wernerSetsTypesTypes1997}).

An advantage of developing type-theoretic constructs on top of an existing set theoretic foundation is that it allows us to further extend type-theoretic features while guaranteeing soundness. As an example, our implementation enriches our dependent function-type interface with a \textit{proof-of-concept} form of subtyping, supporting subtyping assumptions as well as function types covariant in the codomain.

\newcommand{\smartparagraph}[1]{\smallskip \noindent \textbf{#1}. }

\smartparagraph{Contributions} In this paper, we present a set-theoretic realization of dependent types within Lisa. We provide an interface in Lisa that offers syntax and automated reasoning for dependent types with universes, and we implement proof tactics corresponding to evaluation and type checking, based on bidirectional type checking. Additionally, our proof-producing type checker supports a form of subtyping, serving as a proof-of-concept toward the design of a richer set-theoretic type system. Our interface is an extension of Lisa that is fully compatible with both past and future developments, with types represented as set membership sub-formulas. We demonstrate practical use cases of our extension in conjunction with Lisa's library. This approach enables the automated derivation of set-theoretic proofs corresponding to dependently typed judgments. 
We view our work as a first step towards translating statements and proofs from Lean~4 and Rocq into set-theoretic proof assistants \cite{brown_et_al:LIPIcs.ITP.2019.9, 10.1007/978-3-540-69407-6_52, guilloud2023lisa}. 

\section{Embedding of Dependent Product}

In first-order logic, variable binding occurs only through the quantifiers $\forall$ and $\exists$, but embedding type theory requires representing \textit{abstractions} $\lambda x. e$, i.e. \textit{terms} binding variables. Though it is still possible to embed such $\lambda$-terms, it requires complex and inefficient encoding \cite{guilloud2024mechanized}. In this paper we make use of recently developed $\lambda$FOL within Lisa, an extension of first-order logic with $\lambda$-terms and Hilbert's epsilon operator $\varepsilon$. In $\lambda$FOL, expressions are given by simply typed $\lambda$-terms over two primitive sorts: \texttt{Prop} (propositions) and \texttt{Ind} (individuals). Over these atomic sorts, predicates and functions are typed using an arrow constructor, as is customary in systems based on higher-order logic.\footnote{In Lisa, arrow types are written as \texttt{>>:}, which is right-associative.} For instance, membership \(\in\) is a binary predicate with type \(\texttt{Ind} \to \texttt{Ind} \to \texttt{Prop}\) and the quantifier \(\forall\) is a constant of type \((\texttt{Ind} \to \texttt{Prop})\to \texttt{Prop}\). $\varepsilon$ is a quantifier of type \((\texttt{Ind} \to \texttt{Prop})\to \texttt{Ind}\) so that $\varepsilon x. P(x)$ represents some value satisfying $P$, if one exists. Crucially, the only rules given by Lisa to reason with these $\lambda$-terms are $\beta$-reduction and instantiation of schematic symbols (on top of the rules of classical sequent calculus). Hence it is in principle always possible to unfold higher-order terms and definition into pure first-order terms, justifying that the system is still, in spirit, first-order.\footnote{For example, it is impossible to define  equality or universal quantification on higher-order types.} Using $\lambda$FOL and set-theoretic axioms, we embed terms from dependent type theory as follows.
We use $L$ to range over $\lambda$FOL terms of sort
$\texttt{Ind} \to \texttt{Ind}$.

\begin{definition}
Define the embedding $t \mapsto s$ of terms $t$ (following the syntax of pure type systems) into $\lambda$FOL terms. Variables and constants of type theory are represented by variables and constants of FOL and the remaining mappings are as follows:
$$
\begin{array}{rcl}
\lambda x : T .\, u & \mapsto & \abs(T)(\lambda x. u) \\
t\,u & \mapsto & \app(t)(u) \\
\Pi x : T_1 .\, T_2 & \mapsto & \Pii(T_1)(\lambda x. T_2)
\end{array}
$$
where
$$
\begin{array}{rcl}
\abs(T)(L) & := & \{ \langle x, L(x) \rangle \mid x \in T \}\\
\app(t)(u) & := & \varepsilon y.\,(u, y) \in t\\
\Pii(T_1)(L) & := & \left\{
  f \in \mathcal{P}\!\left(T_1 \times \bigcup_{x \in T_1} L(x)\right)
  \;\middle|\;
  \text{isFunc}(f) \land \forall x \in T_1.\, \app(f)(x) \in L(x)
\right\}
\end{array}
$$
\end{definition}
We define $\abs$, $\app$ and $\Pi$ in Lisa once and for all and
prove their necessary properties, so that the particular definition can then be abstracted away. Note that the untyped lambda terms $\lambda x.u$ and $\lambda x.T_2$ refer to the $\lambda$FOL abstraction, 
whereas $\text{isFunc}(f)$ abbreviates the condition $\forall x \in T_1.\, \exists!\,y.\,\langle x,y\rangle \in f$, asserting that $f$ is a functional relation with domain $T_1$.

In type theory, abstraction together with application gives rise to computation via $\beta$-reduction, where a given argument is substituted into the body in a single reduction step. Accordingly, we show that our  definitions of $\abs$ and $\app$ recover the expected $\beta$-reduction behaviour, using $\lambda$FOL to state schematic theorems.

\begin{theorem} [$\beta$-reduction]
If $t \in T$ then $\app(\abs(T)(L))(t) = L(t)$.
\end{theorem}
Based on this definition, we can also prove theorems in set theory corresponding to the inference rules from dependent type theory.
\begin{theorem}[$T_{abs}$]
\label{thm:T_abs}
If $\forall x \in T_1.\, u \in L(x)$
then $\abs(T_1)(L) \in \Pii(T_1)(L)$.
\end{theorem}

\begin{theorem}[$T_{app}$]
\label{thm:T_app}
If $f \in \Pii(T_1)(L)$ and $u \in T_1$, then 
$\app(f)(u) \in L(u)$.
\end{theorem}

The proofs follow from the definitions of $\abs$, $\app$ and $\Pii$ and from set-theoretic axioms. 
Note that we also have the Conversion rule: if $A = B$ then $t \in A \implies t \in B$. We obtain it for free using the substitution property of equality of $\lambda$FOL.

\section{Universe Management}

The standard axioms of ZFC do not directly provide an internal notion of universes suitable for interpreting the cumulative type hierarchy, since there is no set closed under all set-theoretic constructions. In particular, there is no set closed under the $\Pii$ operator. This stands in tension with the requirement that types themselves inhabit higher universes and prevents \textit{polymorphism}, that is abstraction over types. To model a hierarchy of universe $(\mathrm{Type}_i : \mathrm{Type}_{i+1})$\footnote{We use cumulative universes, as in Rocq.} in set theory, we need Tarski’s axiom, ensuring the existence of infinitely many universes, each closed under ZFC operations (and hence under $\Pii$).

\begin{definition}[Grothendieck Universe]\label{def:Grothendieck}
A set $U$ is called a \emph{Grothendieck universe} if it satisfies the following properties: 
(Transitivity) If $y \in U$, then $y \subseteq U$; 
(Pairing, union, and power-set closure) If $y, z \in U$, then $\{y, z\} \in U$, $\bigcup y \in U$, and $\mathcal{P}(y) \in U$;
(Replacement closure) If $A \in U$ and $G$ is a functional relation on $A$ with values in $U$, then the image of $A$ under $G$ is also an element of $U$.
\end{definition}

\begin{definition}[Tarski's axiom]
For every set $x$, there exists a Grothendieck universe $U$ such that $x \in U$.
\end{definition}

\noindent
With Tarski’s axiom in place, we define a simpler characterization of universes, denoted \texttt{isUniverse}. In particular, since our later development only requires closure under dependent products, it suffices to use the functional form of replacement, i.e.\ closure under images of functions $f:A\to U$. 

\begin{definition}[{\texttt{isUniverse} predicate}]
\[
\begin{aligned}
\texttt{isUniverse}(U) \;:=\;&
\bigl(\forall x \in U.\; x \subseteq U \;\land\; \bigcup x \in U \;\land\; \mathcal{P}(x) \in U\bigr) \;\land \\
&\bigl(\forall x \in U.\; \forall y \in U.\; \{x,y\} \in U\bigr) \;\land \\
&\bigl(\forall A \in U.\; \forall f \in U.\; (f : A \to U \Rightarrow \mathrm{range}(f) \in U)\bigr).
\end{aligned}
\]
\end{definition}
Here $A \to U $ is defined as usual as the non-dependent set of functions with domain $a$ and codomain $U$.
Every Grothendieck universe satisfies \texttt{isUniverse}, and this restriction simplifies subsequent automated reasoning.
\begin{lemma}[From Grothendieck universes to \texttt{isUniverse}]
\label{lem:GrothendieckImpliesIsUniverse}
If $U$ is a Grothendieck universe (Definition~\ref{def:Grothendieck}), then $\texttt{isUniverse}(U)$ holds.
\end{lemma}
Let $\universeOf(x)$ denote a universe containing $x$: 
\[ 
\universeOf(x) := \varepsilon U.\, (x \in U \land \texttt{isUniverse}(U)).
\]
Note that we need not require this universe to be minimal.
A hierarchy of infinite universe arises from iterative application of $\universeOf$: $Type_n := \universeOf^n(\emptyset)$. The hierarchy of universe can also be expressed relative to the atoms in the types, such as variables or expressions whose universe are not precisely known.
From these definitions, we can prove key lemmas leading us to the universe-related inference rules $T_{\mathrm{sort}}$ and $T_{\mathrm{form}}$.

\begin{lemma}[Universe closure under dependent products]
\label{lem:universePiClosure}
Let $U$ be a universe. If $T_1 \in U$ and $\forall x \in T_1.\, T_2(x) \in U$, then $\Pi(x:T_1).T_2(x) \in U$.
\end{lemma}
This follows from closure of universes under the operations used to define $\Pii$. 

\begin{theorem}[$T_{\mathrm{sort}}$]\label{thm:T_sort}
For every universe $U$, we have $U \in \universeOf(U)$.
\end{theorem}

\begin{theorem}[$T_{\mathrm{form}}$]\label{thm:T_form}
Let $U_1,U_2$ be universes. Then there exists a universe (denoted $U_1 \sqcup U_2$) such that if $T_1 \in U_1$ and $\forall x. x \in T_1 \implies T_2(x) \in U_2$, 
\[
\Pi(x:T_1).\,T_2(x) \in (U_1 \sqcup U_2).
\]
Moreover, $\universeOf^m(x) \sqcup \universeOf^n(x) = \universeOf^{\max(m, n)}(x)$
\end{theorem}
Proving that the class of all universes is totally ordered by the subset relation $\subseteq$\footnote{this holds, since universes correspond to large cardinals and those are totally ordered} would let us simply define $\sqcup := \cup$, but to simulate countably many universes we actually only need $\sqcup$ at the meta-level.

\section{Proof Generation}

Constructing proofs for concrete typing judgments manually is tedious, so we implement a fully automated proof-generation tactic, \texttt{Typecheck.prove}. The design of this tactic follows bidirectional type-checking discipline \cite{10.1145/3450952}; however, instead of returning a Boolean (or inferred type) as a result, it synthesizes a proof that can be checked by the Lisa kernel.

The proof-generation procedure is organized into three mutually recursive routines, corresponding to the bidirectional structure:
\begin{alphaenumerate}
  \item \textbf{Check mode.}
  Given a term $e$ and an expected type $T_{\mathrm{exp}}$, the routine first inspects whether $e$ is an abstraction.
  If so, $T_{\mathrm{exp}}$ must be a dependent product type. After confirming that the domains coincide, it recursively checks that the body is well-typed, assuming that the bound variable is in the domain, and concludes using $T_{\mathrm{abs}}$ (Theorem~\ref{thm:T_abs}).
  Otherwise, the routine invokes \texttt{infer} mode to compute an inferred type $T_{\mathrm{inf}}$, and then attempts to construct a subtyping judgment $T_{\mathrm{inf}} \subseteq T_{\mathrm{exp}}$ using the conversion rule and the context.

  \item \textbf{Infer mode.}
  Given a term $e$, if it is an application of the form $\app(e_1)(e_2)$, the type of $e_1$ is inferred first and checked to be a dependent product type.
  The argument $e_2$ is then checked against the domain using \texttt{check} mode, and the resulting type is obtained by applying the rule $T_{\mathrm{app}}$ (Theorem~\ref{thm:T_app}).

  If the term is an abstraction, the bound variable is added to the context and the type of the body is inferred; the overall type is then derived using $T_{\mathrm{abs}}$ (Theorem~\ref{thm:T_abs}).

  If the term itself is a dependent product type $\Pi(x:T_1).T_2(x)$, we recursively determine the universe levels of $T_1$ and $T_2(x)$. The resulting ambient universe is then chosen at the meta level as the larger of the two, and the formation rule $T_{\mathrm{form}}$ (Theorem~\ref{thm:T_form}) is applied.

  Otherwise (if the term is a variable, a constant, or any other set-theoretic expression), the routine looks for a possible type in the given context. As a last resort, it infers the type $universeOf(e)$

  \item \textbf{Subset mode.}
  Subset judgments are constructed either directly from assumptions in the context, by reflexivity when the two types coincide, using cumulativity of universes or covariance of function types.
  As a result, the accumulation of subset constraints can be handled entirely by the tactic, rather than requiring explicit user guidance. 
\end{alphaenumerate}

Unlike covariance, contravariance of function types does not carry over directly to a set-theoretic interpretation, as it is a form of impredicativity. Intuitively, for two function types $A \to B$ and $A' \to B'$, subtyping as in many programming languages should yield that if $A' <: A$ and $B <: B'$, $A \to B <: A' \to B'$. However, this implies that the type $A \to B$ should contain \textit{all functions with codomain larger than $A$}: such a set is too big and cannot exist. Hence, we prove the following, more restricted, $\Pi$-subtyping rule:

\begin{lemma}[$\Pi$-subtyping (invariant domain, covariant codomain)]
\label{lem:pi-subtyping}
If $T_1 = T_1'$ and $\forall x \in T_1.\, T_2(x) \subseteq T_2'(x)$, then $\Pi(x:T_1)\,T_2(x) \subseteq \Pi(x:T_1')\,T_2'(x)$.
\end{lemma}
We are exploring avenues to generalize subtyping conditions specifically, as well as the type checking algorithm more broadly.


\section{Example: Polymorphic Composition}
Consider a polymorphic typing judgment corresponding to function composition and the identity function, as it may arise during typechecking in Lean.
\begin{lstlisting}[language=Lean]
def Function.comp{α : Sort u} {β : Sort v} {δ : Sort w} 
  (f : β → δ) (g : α → β) : α → δ
def id{α : Sort u} (a : α) : α
\end{lstlisting}
The example shows using the \texttt{comp} operator to compose a function $f : A \to B$ with a polymorphic identity function 
instantiated to $B \to B$, obtaining a new function $A \to B$.
\begin{lstlisting}[language=scala]
val Typ = variable[Ind]
val Typ3 = getUniverse(3, Typ) 
val Typ5 = getUniverse(5, Typ)

// comp : Π(A B D: Typ_5). (B → D) → (A → B) → A → D
val comp = fun(A :: Typ5, fun(B :: Typ5, fun(D :: Typ5, 
  fun(g :: (B ->: D), fun(f :: (A ->: B), fun(a :: A, g(f(a))))))))

// id : Π(X: Typ_5). X → X
val Id = fun(X :: Typ5, fun(x :: X, x))

val function_id_comp = Theorem((isUniverse(Typ), A ∈ Typ, B ∈ Typ3, 
  f ∈ (A ->: B))  |-  comp(A, B, B, Id(B), f) ∈ (A ->: B)
) { have(thesis) by Typecheck.prove }
\end{lstlisting}
Note that \verb|fun(x :: Typ, e)| corresponds to the abstraction operator 
$\abs(\texttt{Typ})(\lambda x. e)$
introduced earlier, and \verb|A ->: B| denotes the dependent product $\Pi(x : A).\,B$ when $B$ does not depend on $x$. 

In our embedding, universe levels are specified explicitly. Here, we place the polymorphic operators \texttt{comp} and \texttt{id} at universe level $\mathit{Typ}_5$ relative to an arbitrary starting universe \texttt{Typ}. By cumulativity, they can be instantiated with any type living in a lower universe. 

Our tactic, \texttt{Typecheck.prove}, automatically derives set membership corresponding to the typing judgment. The judgment states that, given $A : \mathit{Typ}$, $B : \mathit{Typ}_3$, and $f : A \to B$, composing $f$ with the polymorphic identity function instantiated to $B$ is well-typed as a function from $A$ to $B$. 
Only the base assumption $\texttt{isUniverse}(\mathit{Typ})$ is required, since higher universe predicates follow from Lemma~\ref{lem:GrothendieckImpliesIsUniverse}. Additional examples are provided in the supplementary material.

\section{Conclusion}
We presented a mechanized set-theoretic embedding of dependent products with universes in Lisa, a proof assistant whose kernel is based on first-order logic and set theory. To interpret a cumulative universe hierarchy sufficient for dependent products, we used Tarski's axiom for universes and derived the closure properties needed for type formation. Using this foundation, we implemented a proof-producing bidirectional type-checking tactic that automatically constructs kernel-checkable proof objects for typing judgments. 

\bibliography{ref, sguilloud}

@article{barras1997coq,
  author = {Barras, Bruno and Boutin, Samuel and Cornes, Cristina and Courant, Judicaël and Filliâtre, Jean-Christophe and Giménez, Eduardo and Herbelin, Hugo and Huet, Gérard and Muñoz, César and Murthy, Chetan and Parent-vigouroux, Catherine and Paulin-Mohring, Christine and Saïbi, Amokrane and Werner, Benjamin},
  year = {1997},
  month = {06},
  pages = {},
  title = {The Coq Proof Assistant Reference Manual : Version 6.1}
}

@InProceedings{moura2021lean,
  author="Moura, Leonardo de
  and Ullrich, Sebastian",
  editor="Platzer, Andr{\'e}
  and Sutcliffe, Geoff",
  title="The Lean 4 Theorem Prover and Programming Language",
  booktitle="Automated Deduction -- CADE 28",
  year="2021",
  publisher="Springer International Publishing",
  address="Cham",
  pages="625--635",
  isbn="978-3-030-79876-5"
}

@InProceedings{guilloud2023lisa,
  author =	{Guilloud, Simon and Gambhir, Sankalp and Kun\v{c}ak, Viktor},
  title =	{{LISA - A Modern Proof System}},
  booktitle =	{14th International Conference on Interactive Theorem Proving (ITP 2023)},
  pages =	{17:1--17:19},
  series =	{Leibniz International Proceedings in Informatics (LIPIcs)},
  ISBN =	{978-3-95977-284-6},
  ISSN =	{1868-8969},
  year =	{2023},
  volume =	{268},
  editor =	{Naumowicz, Adam and Thiemann, Ren\'{e}},
  publisher =	{Schloss Dagstuhl -- Leibniz-Zentrum f{\"u}r Informatik},
  address =	{Dagstuhl, Germany},
  URL =		{https://drops.dagstuhl.de/entities/document/10.4230/LIPIcs.ITP.2023.17},
  doi =		{10.4230/LIPIcs.ITP.2023.17}
}

@InProceedings{guilloud2024mechanized,
  author =	{Guilloud, Simon and Gambhir, Sankalp and Gilot, Andrea and Kun\v{c}ak, Viktor},
  title =	{{Mechanized HOL Reasoning in Set Theory}},
  booktitle =	{15th International Conference on Interactive Theorem Proving (ITP 2024)},
  pages =	{18:1--18:18},
  series =	{Leibniz International Proceedings in Informatics (LIPIcs)},
  ISBN =	{978-3-95977-337-9},
  ISSN =	{1868-8969},
  year =	{2024},
  volume =	{309},
  editor =	{Bertot, Yves and Kutsia, Temur and Norrish, Michael},
  publisher =	{Schloss Dagstuhl -- Leibniz-Zentrum f{\"u}r Informatik},
  address =	{Dagstuhl, Germany},
  URL =		{https://drops.dagstuhl.de/entities/document/10.4230/LIPIcs.ITP.2024.18},
  URN =		{urn:nbn:de:0030-drops-207464},
  doi =		{10.4230/LIPIcs.ITP.2024.18}
}

@misc{carneiro2025lean4leanverifyingtypecheckerlean,
  title={Lean4Lean: Verifying a Typechecker for Lean, in Lean}, 
  author={Mario Carneiro},
  year={2025},
  eprint={2403.14064},
  archivePrefix={arXiv},
  primaryClass={cs.PL},
  url={https://arxiv.org/abs/2403.14064}, 
}

@article{10.1145/3450952,
author = {Dunfield, Jana and Krishnaswami, Neel},
title = {Bidirectional Typing},
year = {2021},
issue_date = {June 2022},
publisher = {Association for Computing Machinery},
address = {New York, NY, USA},
volume = {54},
number = {5},
issn = {0360-0300},
url = {https://doi.org/10.1145/3450952},
doi = {10.1145/3450952},
journal = {ACM Comput. Surv.},
month = may,
articleno = {98},
numpages = {38}
}

@InProceedings{10.1007/978-3-540-74591-4_28,
author="Wiedijk, Freek",
editor="Schneider, Klaus
and Brandt, Jens",
title="Mizar's Soft Type System",
booktitle="Theorem Proving in Higher Order Logics",
year="2007",
publisher="Springer Berlin Heidelberg",
address="Berlin, Heidelberg",
pages="383--399",
isbn="978-3-540-74591-4"
}

@misc{urban2026130klinesformaltopology,
      title={130k Lines of Formal Topology in Two Weeks: Simple and Cheap Autoformalization for Everyone?}, 
      author={Josef Urban},
      year={2026},
      eprint={2601.03298},
      archivePrefix={arXiv},
      primaryClass={cs.LO},
      url={https://arxiv.org/abs/2601.03298}, 
}

@InProceedings{brown_et_al:LIPIcs.ITP.2019.9,
  author =	{Brown, Chad E. and Kaliszyk, Cezary and P\k{a}k, Karol},
  title =	{{Higher-Order Tarski Grothendieck as a Foundation for Formal Proof}},
  booktitle =	{10th International Conference on Interactive Theorem Proving (ITP 2019)},
  pages =	{9:1--9:16},
  series =	{Leibniz International Proceedings in Informatics (LIPIcs)},
  ISBN =	{978-3-95977-122-1},
  ISSN =	{1868-8969},
  year =	{2019},
  volume =	{141},
  editor =	{Harrison, John and O'Leary, John and Tolmach, Andrew},
  publisher =	{Schloss Dagstuhl -- Leibniz-Zentrum f{\"u}r Informatik},
  address =	{Dagstuhl, Germany},
  URL =		{https://drops.dagstuhl.de/entities/document/10.4230/LIPIcs.ITP.2019.9},
  URN =		{urn:nbn:de:0030-drops-110643},
  doi =		{10.4230/LIPIcs.ITP.2019.9}
}

@InProceedings{10.1007/978-3-540-69407-6_52,
author="Paulson, Lawrence C.",
editor="Beckmann, Arnold
and Dimitracopoulos, Costas
and L{\"o}we, Benedikt",
title="The Relative Consistency of the Axiom of Choice --- Mechanized Using Isabelle/ZF",
booktitle="Logic and Theory of Algorithms",
year="2008",
publisher="Springer Berlin Heidelberg",
address="Berlin, Heidelberg",
pages="486--490",
isbn="978-3-540-69407-6"
}

@article{coquandCalculusConstructions1988,
  title = {The Calculus of Constructions},
  author = {Coquand, Thierry and Huet, G{\'e}rard},
  year = 1988,
  month = feb,
  journal = {Information and Computation},
  volume = {76},
  number = {2},
  pages = {95--120},
  issn = {0890-5401},
  doi = {10.1016/0890-5401(88)90005-3},
  url = {https://www.sciencedirect.com/science/article/pii/0890540188900053},
  urldate = {2021-06-11},
  langid = {english}
}

@book{jechSetTheory1978,
  title = {Set Theory},
  author = {Jech, Thomas J.},
  year = 1978,
  series = {Pure and Applied Mathematics, a Series of Monographs and Textbooks},
  number = {79},
  publisher = {Academic Press},
  address = {New York},
  isbn = {978-0-12-381950-5},
  langid = {english},
  lccn = {QA3 QA248 .P8 vol. 79}
}

@incollection{wernerSetsTypesTypes1997,
  title = {Sets in Types, Types in Sets},
  booktitle = {Theoretical {{Aspects}} of {{Computer Software}}},
  author = {Werner, Benjamin},
  editor = {Goos, Gerhard and Hartmanis, Juris and {van Leeuwen}, Jan and Abadi, Mart{\'i}n and Ito, Takayasu},
  year = 1997,
  volume = {1281},
  pages = {530--546},
  publisher = {Springer Berlin Heidelberg},
  address = {Berlin, Heidelberg},
  doi = {10.1007/BFb0014566},
  url = {http://link.springer.com/10.1007/BFb0014566},
  urldate = {2021-05-14},
  isbn = {978-3-540-63388-4 978-3-540-69530-1},
  langid = {english}
}

\end{document}